# On the Order of Gene Distribution on Chromosomes Across the Animal Kingdom

Abdullah A. Toor, [1] Amir A. Toor MD. [2]

School of World Studies,[1] and Department of Internal Medicine,[2] Virginia Commonwealth University, Richmond, VA 23298.

Address correspondence to, Amir A. Toor MD,

Professor of Medicine, Department of Internal Medicine,

Massey Cancer Center, Virginia Commonwealth University,

Richmond, VA 23298. Phone: 804-628-2389, Email: amir.toor@vcuhealth.org

Summary

**Background.** The large-scale pattern of distribution of genes on the chromosomes in the known animal genomes is not well characterized. We hypothesized that individual genes will be distributed on chromosomes in a mathematically ordered manner across the animal kingdom. **Results.** Twenty-one animal genomes reported in the NCBI database were examined. Numerically, there was a trend towards increasing overall gene content with increasing size of the genome as reflected by the chromosomal complement. Gene frequency on individual chromosomes in each animal genome was analyzed and demonstrated uniformity of proportions within each animal with respect to both average gene frequency on individual chromosomes and gene distribution across the unique genomes. Further, average gene distribution across animal species followed a relationship whereby it was, approximately, inversely proportional to the square root of the number of chromosomes in the unique animal genomes, consistent with the notion that there is an ordered increase in gene dispersion as the complexity of the genome increased. To further corroborate these findings a derived measure, termed gene spacing on chromosomes correlated with gene frequency and gene distribution. **Conclusion.** As animal species have evolved, the distribution of their genes on individual chromosomes and within their genomes, when viewed on a large scale is not random, but follows a mathematically ordered process, such that as the complexity of the organism increases, the genes become less densely distributed on the chromosomes and more dispersed across the genome.

## 1. Introduction.

The animal kingdom is characterized by incredible diversity among thousands of species, each demonstrating unique characteristics that distinguish them from others. The unique morphology of each animal has its origin in the genetic variation encountered between species, and has taken billions of years to evolve to its current level of diversity. The genetic material in eukaryotic organisms is organized into protein coding regions – genes – interspersed with vast tracts of noncoding DNA, which has recently been shown to be extensively involved in regulation of gene expression. [1, 2, 3, 4] Millions of base pairs of DNA comprising these genes are arranged on histone-DNA aggregates, the chromosomes [5]. The animal kingdom demonstrates a wide variety of genetic configuration with various animal species possessing different number of genes distributed upon a heterogeneous background of varying chromosome numbers. The numbers of genes on each chromosome varies between several hundred to several thousand, with each animal possessing different number of chromosomes. Aside from the knowledge that more complex animals possess a larger number of genes and chromosome complement than smaller animals, there is no known large-scale quantitative relationship describing the relative distribution of genes on different chromosomes across the animal kingdom.

A challenge in this respect has been the scaling variation encountered in the way genetic information is coded across the animal kingdom. Animals have a variable number of chromosomes, with information coding for a large number of proteins (genes), found in sequences of nucleotides, which are much larger still. This makes finding a common organizational framework difficult. Recently models based in non-Euclidean geometry, such as fractals, have been utilized to study this problem. Objects in nature tend to have fractal geometry, and similar organization has been observed in the molecular folding of DNA, and in nucleotide distribution in the genome [6, 7, 8, 9]. Fractal organization describes similarity in patterning across scales of magnitude, and is a structural motif often observed in nature, such as in the branching patterns of trees, or in the arborization of vascular and neuronal networks in animals [10, 11, 12, 13]. These observations have been extended to the human T cell repertoire, demonstrating fractal patterning in terms of T cell receptor β, variable, joining and diversity segment usage in forming the repertoire [14].

Given the widespread presence of self-similar, fractal structures in nature, similarity of patterning and proportion in terms of information coding between various animal genomes was investigated, with the hypothesis that information (genes) will be coded on chromosomes in a self-similar, fractal distribution. Further this patterning may be similar across animals with varying genetic complements, conforming to an underlying pattern rather than each animal having randomly assorted sets of genes. To accomplish this a simplifying assumption was made, that is, individual genes constitute units of information and their distribution on chromosomes may be viewed conceptually as, beads on a string, or more formally, as prime numbers distributed among a sequence of natural numbers. In this paper the results of this quantitative evaluation of gene distribution across chromosomes and genomes of several animals is reported. Uniformity across animal genomes with proportional distribution of genes on chromosomes is evident when logarithms are used to overcome the effect of varying scales. Our findings suggest that there is an underlying mathematical order to the distribution of genes on chromosomes in the animal kingdom, which may be a consequence of a mathematically determined process of evolution.

## 2. Methods.

2.1. To investigate the distribution of genes on individual chromosomes, data on animals where the entire genome is characterized and recorded in the National Center for Biotechnology Information website [15, 16] were obtained in July 2012. A total of 21 animals were examined with a chromosome complement ranging from 3 to 39 chromosomes with a variable number of genes on each chromosome and varying chromosome sizes, measured in million base pairs (Mbp) (Table 1). The primary data comprised of the number of chromosomes, the number of genes, and chromosome size in number of base pairs. Data was arranged in individual tables for each animal with the size of each chromosome, aligned with its gene content. Numeric data were transformed to base 10 logarithms to eliminate the effect of different scale of magnitude between the variables being examined, and to allow comparisons across different scales, e.g. chromosome size in millions of base pairs, versus gene number on each chromosome in the thousands, and chromosome numbers in each animal in tens. Gene frequency was compared across all the chromosomes in individual animals and between animals. Regression analyses were performed using central values of various derived functions identified in the course of the computations undertaken, comparing them between different animal genomes.

2.2. Gene distribution on individual chromosomes was examined, hypothesizing that it would exhibit self-similar fractal organization. Self-similarity is generally used to describe scale invariance of magnitude, maintaining a proportional relationship between the magnitude of a variable and the scale of measurement. This may be revealed by computing the ratio of logarithm of magnitude and logarithm of the scale of measurement, which should retain a relatively constant value. [17] To explore the fractal nature of gene distribution within and across individual genomes, it was assumed that the gene distribution from chromosome to chromosome in an animal would be similar, and that these values from individual genomes, when compared across the animal kingdom would also exhibit proportionality, characteristic of fractal organization.

2.2.1. Frequency of occurrence of genes on each chromosome was determined by comparing the number of genes on individual chromosomes with their sizes (in base pairs of DNA) within an animal, using the formula

$Gene\ frequency$ for chromosome $_i$ = Log $(x_i)$ / Log $(m_i)$ …….. [1]

where $x_i$ is the number of genes on the $i$ th chromosome and $m_i$ is the size of chromosome in base pairs of DNA for that chromosome. In this calculation, the size of the chromosomes, given in nucleotides, was assigned a scale value, since it is the template on which the genes (information) are inscribed. Average $gene\ frequency$ for each animal was then determined to account for all the chromosomes in its genome.

2.2.2. To determine self-similarity of gene distribution across the genome of each animal, the haploid number of chromosomes for each animal, $n$, was given a unique scale value for that animal.  The number of genes on the $i^{th}$ chromosome, $x_i$, were then used in the following formula,

$Relative\ gene\ distribution$ on chromosome $_i$ = Log $(x_i)$ / Log $(n)$……… [2]

For each animal the average value for all the chromosomes was calculated and this was termed average $relative\ gene\ distribution$ function as it represented, the pattern of occurrence of genes on each chromosome relative to the entire genome of the animal as reflected by the haploid chromosomal content in individual species. This allows a comparison of how information is coded over the entire genome in different animals.

2.3. Genes occur on each of the chromosomes with a certain periodicity. To estimate the magnitude of this periodicity, and use it to compare similarity between genomes, a comparison of the number of genes on individual chromosomes was made with a fundamental periodic phenomenon encountered in nature, i.e., the occurrence of prime numbers in a sequence of natural numbers. This may be best estimated by calculating the Riemann zeta function for a sequence of numbers ($see\ appendix$). Spacing of the non-trivial zeros of the Riemann zeta function on the critical line may model the distribution of genes on chromosomes and account for the periodicity of their occurrence. The average space between the non-trivial zeros of the Riemann zeta function on the critical line is $\sim 2\pi$ / Log $(T / 2\pi)$. (18) The number of genes on each chromosome, $x_i$ was substituted for *T* in the original formula to obtain the dimensionless *average gene spacing* parameter for specific chromosomes.

*Average gene spacing* on chromosome $i \approx 2\pi \, / \, \mathrm{Log} \, (x_i / 2\pi)$……. [3]

Once again these values for all the chromosome in each animal were averaged for comparison across species.

2.3.1. In all the formulae, noted above it is to be recognized that the relationships are approximate, since genes are not discrete elements, but rather are made up of protein coding *exons* interspersed with non-coding *introns*; therefore these calculated values or functions are estimates rather than exact relationships.

2.4. *Statistical analysis*

Linear associations between various gene measurements were estimated using simple linear regression and by Spearman's rank correlation coefficient (where the coefficient of determination ($R^2$) and estimated correlation (R) are presented along with their corresponding p-values (P)). Non-linear associations between various gene measurements are estimated using polynomial regression or non-linear regression (where the coefficient of determination ($R^2$) is presented along with its corresponding p-value (P)). SAS version 9.2 (Cary, NC, U.S.A.) is used for all statistical analyses, specifically using the REG, CORR and NLIN procedures.

## 3. Results.

### 3.1. *Variation in genome size across animal kingdom*

Initial analyses were conducted to determine simple numeric relationships between the size of each chromosome and its gene content. A linear positive relationship between the total gene complement of animals and the size of the entire genome expressed in mega-base pairs of DNA was observed (Spearman rank correlation coefficient R = 0.69, P<0.001) (Figure 1A). Further, when the total number of genes or DNA content of a genome were examined in relation to its complexity (reflected by chromosome numbers), weak quadratic relationships were observed ($R^2$ = 0.39, P = 0.016 for number of genes vs. number of chromosomes; $R^2$ = 0.39, P = 0.015 for DNA content vs. number of chromosomes) (Figure 1B and 1C). Examining overall trends, the animals tended to form groups demonstrating similar characteristics along the quadratic curve, with invertebrates and avian life forms occupying the areas near the two x-intercepts of the quadratic plot, and mammals – particularly primates – occupying the apex.

### 3.2. *Self-similarity in gene distribution within individual chromosomes*

Gene distribution on individual chromosomes was next examined hypothesizing that it would exhibit uniformity across each animal's genome. Log-transformation of the data was used to eliminate the effect of scale of measurement, and demonstrated similarity in proportion of individual chromosome sizes in base pairs and number of genes present on them, when depicted graphically for individual animals (Figure 2A-D). This uniformity in the pattern of distribution of genes on chromosomes of varying sizes was evident across the animals examined.

When the number of genes present on individual chromosomes was measured, once again, uniformities were observed in the ratios of the logarithm of gene numbers on each chromosome with the logarithm of its size in base pairs, within each animal. The average value was a reflection of the gene frequency per chromosome for that animal's genome (Table 2). The average *gene frequency* remained more or less constant across the animal kingdom, demonstrating a slow decline ($R^2$=0.73, P<0.001) as the complexity of various animals increased, reflected by increasing number of chromosomes (Figure 3). Thus, if one considers the different chromosomal content of each animal, the relative uniformity of the *gene frequency* across all the animals studied, is suggestive of a fractal, self-similar organization of the genes in the genomes across the animal kingdom.

### 3.3. *Similarity in gene distribution within individual genomes*

Given the above relationship, similarity of gene distribution across the genome of each animal was investigated using the haploid chromosome content as the scaling factor. This was termed *relative gene distribution* function and represented, the pattern of occurrence genes on each chromosome relative to the entire genome of the animal. This *gene distribution* function though uniform in each animal, varied across the different animals in the kingdom (Table 2), demonstrating high values consistent with tight gene clustering in less evolved animal such as invertebrates and a more dispersed pattern with lower *gene distribution* function values in more evolved organisms such as mammals and birds. Further, there was positive correspondence between *gene frequency* and *gene distribution function* (Spearman rank correlation coefficient R=0.91; P<0.001). This suggested that gene distribution, or information packaging in the genetic material of each species conforms to an underlying pattern, in other words, genes are not distributed on chromosomes in a random manner but rather are distributed in relation to the number and size of the chromosomes present in each animal.

### 3.4. *Similarity in gene distribution across animal species*

Regression analysis was then performed using central values of the *relative gene distribution* function comparing them between genomes ordered by the scale value utilized, i.e., the number of chromosomes, *n*, across the animal kingdom. The *relative gene distribution* function demonstrated a progressive decline with increasing chromosome content when it was compared across different animal species. Notably, when plotted against the chromosome complement of individual animals it declines as a power function of the haploid chromosomal complement ($R^2$= 0.98, P<.001) (Figure 4A).

Although this is a function of the mathematical operation at hand, an examination of how the species are distributed on this curve demonstrates an evolutionary continuum among the animals studied, extending from the simplest (invertebrates) at the apex, to the most complex genomes (mammals and birds) as the curve approaches the asymptote. Further, the plot of average *relative gene distribution* function against the number of chromosomes demonstrates approximate inverse proportionality of gene distribution to the square root of the haploid chromosomal content, delineating the relationship between genetic information and genetic material across animal species. This is consistent with a mathematically ordered increase in gene dispersion as the chromosomal complement of the genomes increased. The absolute value of the exponent in this limited sample is 0.55, which is close to the value of the Gamma constant (0.57) describing the relationship between the harmonic series and natural logarithm of numbers. This observation was further reinforced by the similar distribution observed when non-log transformed data were considered, i.e. median number of genes per chromosome for each animal, plotted against the haploid chromosomal content of that animal ($R^2 = 0.67$; $P<.001$) (Figure 4B). This observed clustering of animals along the various parts of the regression curve is similar to that observed earlier, with invertebrates occupying the top ranks and primates clustering together in the middle, and animals with larger chromosomal complements grouped together at the far end of the resulting curve.

3.5. *Relative spacing of genes across chromosomes*

It was conjectured that the distribution of genes on chromosomes, might be modeled by the periodic occurrence of the non-trivial zeros of Riemann's zeta function. The derived value was termed *average gene spacing,* and calculated for each chromosome demonstrated a narrow distribution within each animal despite the heterogeneity in the gene content of individual chromosomes. Further, when average gene spacing for each chromosome was plotted against its gene content (Table 3), within animals, it declined as a power function, (Fig. 5B, 5C) implying that the relative gene spacing diminished by about the $1/5^{th}$ - $1/7^{th}$ power of the number of genes on a chromosome, as that value increased. This relationship is analogous to the Riemann zeta-zero distribution, where the zeros become more tightly clustered on the critical line, as they are calculated for increasing values of *y* on the imaginary axis of the complex plane. As in the gene distribution example above, the observed power law distribution of *gene spacing* with gene content on chromosomes in each animal, most likely reflects an inherent property of the transformation, *(2π) /* Log *(x/(2π))* plotted against *x*. However, there is a steady linear increase in the coefficients and exponents across the animal kingdom, as the average *gene spacing* value increases with the rising chromosome complement (Table 3). Further, the increase in the *average gene spacing* with an increasing chromosomal complement is consistent with the earlier observed reduction in *gene distribution* function values and *gene frequency* in more complex animals. The inverse correlation between these measures was confirmed by the linear relationship between the reciprocal of *gene spacing* and average *gene content* calculated for each animal (Figure 6). This implies that as the complexity of the organism increases, as reflected by a larger chromosomal and gene complement, the regulatory, non-coding elements increase, all in a mathematically determined fashion.

## 4. Discussion.

Biologic complexity has evolved over millennia with environmental influences promoting genetic mutations. Mutations imparting a survival advantage are then perpetuated, resulting in the species diversity that the living world is imbued with. Therefore to model evolution at the molecular scale one would need to begin with determining the rules underlying the variation in the genomes of various organisms. However the heterogeneity and abundance of genetic information make this challenging. In this paper an attempt has been made to discern the large-scale pattern of variation across the genome of various animals studied to date. A very simple model has been examined – i.e. studying the distribution of genes, as units of information, on chromosomes, as repositories containing the information, across various species – to get an inkling of the principles underlying genetic variation. We report a degree of similarity in the examined animal species, which demonstrates that the protein coding genes – far from being randomly distributed across the genomes of various organisms – demonstrate constant proportionality to the genetic material comprising each species' genome. Further, across species there is a remarkable degree of similarity in this relationship suggesting that gene distribution may have a fractal ordering.

The organization of recorded genetic information in the individual animal genomes can be discerned by examining the structures, which exist naturally, i.e. genes and chromosomes. In our study, *gene frequency* on each chromosome was measured; genes were considered as units of information, and chromosome length in base pairs, as the coding material on which this information is recorded. When logarithms are used to eliminate the effect of scale of measurement on these quantities in each animal, a striking level of uniformity is observed in the resulting *gene frequency* which extends beyond the species, linking related species by demonstrating clustering of their values, such as in the primates. Thus chromosomes represent not a random aggregation of genetic material, rather structured collection of information. Next we considered how this information was organized across the genome in each animal by examining gene distribution. Once again when logarithms are used to eliminate the effect of scaling between gene number per chromosome and number of chromosomes in the genome, it becomes clear that each animal packages the genetic information more or less uniformly across its genome, and that this packaging is also reproduced in related species. An interesting corollary to this observation is the quadratic relationship observed between total gene content or genome size and the number of chromosomes. Starting from the animals with a smaller number of very tightly packed genes among the invertebrates, the genetic information reaches an optimally 'packaged' apex distribution of a large number of genes in the primates. As the number of chromosomes increases, however the overall genetic information declines in avian and other mammalian life forms as genes become more dispersed and smaller chromosomes with minimal gene complement are encountered.

Finally, to verify the information (gene) assembly principles elucidated above, the number of genes on each chromosome was examined independent of the size of the chromosomes and genomes. This derived parameter, termed *average gene spacing* since it derives from the formula for calculating average spacing of the non-trivial Riemann zeta-zeros, follows the same trends across the individual species as well as the across the animal kingdom, affirming that gene distribution on chromosomes across animal kingdom is highly regulated within species, and between species as well. This demonstrates that a quantitative relationship exists between the genetic information coded and the genetic material on which it is coded. This may in turn be taken to demonstrate that evolution can be viewed as a mathematically ordered process.

The similarity observed in the parameters detailed above, both within animal genomes and between different animals suggests fractal ordering of gene distribution across species in the animal kingdom. Nature abounds with processes that conform to fractal organization. Fractals are mathematical constructs, which – through a process of iteration – generate an array of complex geometrical structures which take on different forms based on the initial conditions, yet are all characterized by similarity of proportions across scales. First described by Benoit Mandelbrot, fractals have been used to explain scale invariance of measurements in a variety of natural and theoretical systems. Similar relationships have been previously described, in which box-counting methodology demonstrated fractal organization of four different genomes including the human genome. [19] Our finding of similarity in gene distribution across the chromosomes in the genomes of various organisms is likewise reminiscent of the self-similar processes at work in generation of fractals. This argument may be applied to the gene frequency function we have described, in which the number of genes present on each chromosome, accounting for the DNA content of each chromosome, maintains a relatively narrow distribution which remains more or less unaltered across chromosomes of different sizes and across different animal species. Likewise, when the distribution of genes across the genome was examined by accounting for the haploid chromosomal content of various animals it was found to maintain fairly narrow bounds within each animal. This is depicted in the proposed model of gene distribution (Fig 7), which demonstrates self-similarity within and across genomes, specifically maintenance of proportionality in the distribution of genes across the chromosomes in various genomes. Recent findings of the ENCODE project give further credence to our findings of increasing gene dispersion with greater complexity of the organism's genome. Intergenic DNA is in turn likely involved in the increasingly complex regulatory function in the more evolved organisms.

Importantly, comparing the gene distribution across animals revealed that when ordered by chromosome content (or genome complexity) gene distribution follows a distribution, which demonstrates that in a given animal the *average gene distribution* is inversely proportional to the square root of the haploid chromosome content of that animal. In fact the exponent in this relationship approaches the well-known $\gamma$-constant of Euler and Mascheroni, which describes the relationship between the harmonic series and natural logarithm of a number. [20] This and the observation that log-transformed quantitative relationships describing gene content of various chromosomes followed principles comparable to complex number iteration, that is, self-similar fractals, it was hypothesized that gene distribution in chromosomes might be modeled in a fashion analogous to the distribution of the non-trivial zeros of Riemann's $\zeta$ function, as was the case. The mathematically ordered configurations taken on by different genomes, evokes the notion of the achievement of stable quantum states being important in terms of DNA assembly and thus gene distribution across genomes. Stable nuclear energy states as well as energetics at the quantum-classical boundary are described by the same relationships that define the spacing of Riemann $\zeta$-zeros on the critical line. Taking the logical position that, as in those distributions, one may hypothesize that frequency of occurrence of genes on chromosomes will follow similar laws of periodic functions. Taking a simple model of chromosomes and genes representing beads on a string, the average spacing of which declines logarithmically and is quantified as a periodic function, with periods taking on values which are fractions of $2\pi$. This dimensionless function, *average gene spacing*, which revealed strikingly similarity between animals regardless of their position on the evolutionary scale. That it follows similar principles as $\zeta$-zero distribution, which is also replicated by the eigenvalues of random matrices as well as the energy states of quantum classical boundary, suggests that evolution may be driven by the same quantum-classical boundary chaos mechanisms which describe those phenomenon. [21] This makes it entirely plausible that the evolutionary process follows similar 'rules' when packaging information on DNA, with a limited number of possibilities leading to stable configuration and thus likely to lead to successful phenotypic changes in the organism resulting in evolutionary advantage. Thus it appears that the genome is organized with an underlying mathematical order analogous to the distribution of primes, and the energy levels at the quantum classical boundary.

Although examined at a relatively 'low-resolution' of information packaging, i.e. gene distribution on chromosomes in different animal species, our findings suggest that evolution – rather than being a random process -- is a highly ordered process that may be determined using mathematical principles, such as iteration in the context of logarithmically determined proportions.  This work can only be taken as a small step in trying to determine the underlying order of distribution of protein coding elements in animal genomes, raises the intriguing possibility that evolution may be studied as an example of a 'chaotic' process. Further work accounting for the complexity of gene 'structure' -- i.e. incorporating nucleotide sequence and more rigorous mathematical analysis of the resulting data – will hopefully further our understanding of the organizational principles introduced here.

## 5. Conclusion.

Large-scale evaluation of gene distribution on chromosomes in animals, demonstrates similarity of proportions between gene content and distribution on individual chromosomes. This self-similarity provides a unique insight into the evolution of species from a genomic perspective, underscoring that evolution is a mathematically ordered process. Aside from suggesting the presence of an underlying order to evolution in the animal kingdom, the associations observed hint at the role of rules involving the quantum-classical boundary phenomenon being important in maintaining genetic stability.

**Table 1.** Animals included in the study with the NCBI reference data set as of July 2012.

| Animal | Conventional nomenclature | Reference Data set | Number of chromosomes* |
|---|---|---|---|
| *Anopheles gambiae* | African malaria mosquito | AgamP3.3 | 3 |
| *Nasonia vitripennis* | Jewel wasp | Build 2.1 | 5 |
| *Caenorhabditis briggsae* | Nematode | Build 1.1 | 6 |
| *Anolis carolinensis* | Green anole | Build 1.1 | 6** |
| *Monodelphis domestica* | Opossum | Build 2.2 | 9 |
| *Sus scrofa* | Pig | Sscrofa10.2 | 19 |
| *Mus musculus* | Laboratory mouse | Build 38.1 | 20 |
| *Macaca mulatta* | Rhesus macaque | Build 1.2 | 21 |
| *Rattus norvegicus* | Rat | RGSC v3.4 | 21 |
| *Oryctolagus cuniculus* | Rabbit | OryCun2.0 | 22 |
| ***Homo sapiens*** | **Human** | **Build 37.3** | **23** |
| *Callithrix jacchus* | White-tufted-ear marmoset | Build 1.1 | 23 |
| *Pan troglodytes* | Chimpanzee | Build 2.1 | 24 |
| *Pongo abelii* | Sumatran orangutan | Build 1.2 | 24 |
| *Danio rerio* | Zebrafish | Zv9 | 25 |
| *Gallus gallus* | Chicken | Build 3.1 | 29 |
| *Bos taurus* | Cattle | Build 6.1 | 30 |
| *Meleagris gallopavo* | Turkey | Build 1.1 | 31 |
| *Equus caballus* | Horse | EquCab2.0 | 32 |
| *Taeniopygia guttata* | Zebra finch | Build 1.1 | 32 |
| *Canis lupus familiaris* | Dog | Build 3.1 | 39 |

* Haploid chromosome content; ** 6 chromosomes & 7 linkage groups

[21] NCBI: http://www.ncbi.nlm.nih.gov/projects/mapview/

**Table 2.** Mean (±standard deviation) *gene frequency* (log gene number on chromosome/log chromosome size in bp; formula [1]) and *gene distribution* function (log gene number on chromosome/log haploid chromosome number, formula [2]) for individual species. Note clustering of values for related species such as primates.

| Animal | Gene frequency | | Gene distribution | |
|---|---|---|---|---|
| Nematode | 0.48 | 0.01 | 4.41 | 0.10 |
| Mosquito | 0.45 | 0.03 | 7.36 | 0.87 |
| Jewel wasp | 0.44 | 0.00 | 4.81 | 0.07 |
| Zebra fish | 0.40 | 0.02 | 2.23 | 0.14 |
| Mouse | 0.39 | 0.02 | 2.46 | 0.15 |
| Human | 0.39 | 0.02 | 2.30 | 0.17 |
| Opossum | 0.39 | 0.02 | 3.47 | 0.32 |
| Rhesus macaque | 0.38 | 0.02 | 2.36 | 0.13 |
| Rat | 0.38 | 0.02 | 2.34 | 0.15 |
| Orangutan | 0.38 | 0.02 | 2.21 | 0.15 |
| Marmoset | 0.37 | 0.04 | 2.19 | 0.29 |
| Chimpanzee | 0.37 | 0.03 | 2.17 | 0.19 |
| Cattle | 0.37 | 0.02 | 1.97 | 0.14 |
| Pig | 0.37 | 0.05 | 2.31 | 0.36 |
| Chicken | 0.36 | 0.05 | 1.76 | 0.31 |
| Rabbit | 0.35 | 0.02 | 2.10 | 0.21 |
| Dog | 0.35 | 0.02 | 1.73 | 0.14 |
| Horse | 0.35 | 0.03 | 1.84 | 0.18 |
| Turkey | 0.34 | 0.03 | 1.66 | 0.25 |
| Zebra finch | 0.34 | 0.06 | 1.62 | 0.33 |

**Table 3.** Average (±standard deviation) *gene spacing* per chromosome (formula [3]) across the animal kingdom, with power law expression correlating the gene spacing for each chromosome with its gene frequency (*x*).

| Animal | Chromosome N | Avg. gene spacing | | gene spacing/ gene N/chr |
|---|---|---|---|---|
| Mosquito | 3 | 2.35 | 0.39 | $9.5\,x^{-0.17}$ |
| Wasp | 5 | 2.45 | 0.05 | $9.1\,x^{-0.17}$ |
| Nematode | 6 | 2.39 | 0.05 | $8.8\,x^{-0.16}$ |
| Opossum | 9 | 2.55 | 0.33 | $10.0\,x^{-0.18}$ |
| Pig | 19 | 2.87 | 0.31 | $11.3\,x^{-0.20}$ |
| Mouse | 20 | 2.63 | 0.2 | $10.9\,x^{-0.19}$ |
| Macaque | 21 | 2.72 | 0.22 | $10.9\,x^{-0.19}$ |
| Rat | 21 | 2.75 | 0.23 | $10.3\,x^{-0.10}$ |
| Rabbit | 22 | 3.17 | 0.49 | $13.7\,x^{-0.23}$ |
| Humans | 23 | 2.72 | 0.29 | $11.0\,x^{-0.20}$ |
| Marmoset | 23 | 2.81 | 0.29 | $11.5\,x^{-0.21}$ |
| Orangutan | 24 | 2.81 | 0.27 | $11.3\,x^{-0.20}$ |
| Chimpanzee | 24 | 2.9 | 0.4 | $12.9\,x^{-0.22}$ |
| Zebra Fish | 25 | 2.71 | 0.18 | $9.3\,x^{-0.17}$ |
| Chicken | 29 | 3.53 | 0.68 | $15.1\,x^{-0.24}$ |
| Cattle | 30 | 3 | 0.3 | $12.3\,x^{-0.21}$ |
| Turkey | 31 | 4 | 1.4 | $22.0\,x^{-0.31}$ |
| Horse | 32 | 3.26 | 0.49 | $13.9\,x^{-0.23}$ |
| Dog | 39 | 3.27 | 0.38 | $13.6\,x^{-0.23}$ |

Figure Legends

**Figure 1:** Relationship between total number of genes and whole genome size (in bp of DNA or number of chromosomes) in animals. (A) Total number of genes plotted against Mega base pairs of DNA in genomes of different animals. (B) Number of genes plotted against genome complexity, reflected by chromosome number. (C) Genome size in Mbp of DNA plotted against chromosome content.

**Figure 2.** Chromosome size in base pairs (blue area) and gene complement (red area) per chromosome plotted out (log scale) for individual chromosomes in representative animals. Individual chromosome number given on X-axis. (A) *Homo sapiens* (humans); (B) *Gallus gallus* (Chicken); (C) *Anolis carolinensis* (Green anole, 6 chromosomes and 7 linkage groups); (D) *Oryctolagus cuniculus* (Rabbit).

**Figure 3.** *Gene frequency* function [1] plotted against chromosome number in the animals examined (N=20). Similar gene content per chromosome when normalized for chromosome size in all animals, with a slight decline as chromosomal complement of animal increases.

**Figure 4.** Average *gene distribution* across entire genome. (A**)** Median *gene distribution function* [2] plotted against haploid chromosomal content (N=20) (B**)** Median genes per chromosome plotted against haploid chromosomal content (N=20)

**Figure 5.** Power distribution of gene content accounting for chromosome size. (B) *gene spacing* [3] and gene complement of each chromosome for humans. (C) *Gene spacing* ranked in order of magnitude, for individual chromosomes in all the genomes examined, follow Power law relationships. Red squares, average gene spacing per chromosome ($R^2$= 0.58). Each data point corresponds to an individual chromosome.

**Figure 6.** Median *gene frequency* [1] plotted against reciprocal of *gene spacing* [3]. (N=20) Linear increase in gene content per chromosome as the spacing between the genes declines.

**Figure 7.** A model depicting the relationship between gene frequency, gene distribution and gene spacing on three hypothetical genomes A, B and C with increasing chromosomal complement. Two partial chromosomes from each of the genomes are shown. Although the gene content of the chromosomes remains relatively unchanged as evolution proceeds, the *gene frequency* declines, with a proportional decline in *gene distribution* function (genes go from being tightly clustered to more dispersed) and a similar proportional increase occurs in *gene spacing* as animal proceed from simple to more complex organism from left to right. This implies that genes become more disperse in a mathematically ordered fashion as the chromosomal content of the animal increases.

**Figure 1.** Relationship between total number of genes and whole genome size (in bp of DNA or number of chromosomes) in animals. (A) Total number of genes plotted against Mega base pairs of DNA in genomes of different animals. (B) Number of genes plotted against genome complexity, reflected by chromosome number. (C) Genome size in Mbp of DNA plotted against chromosome content.

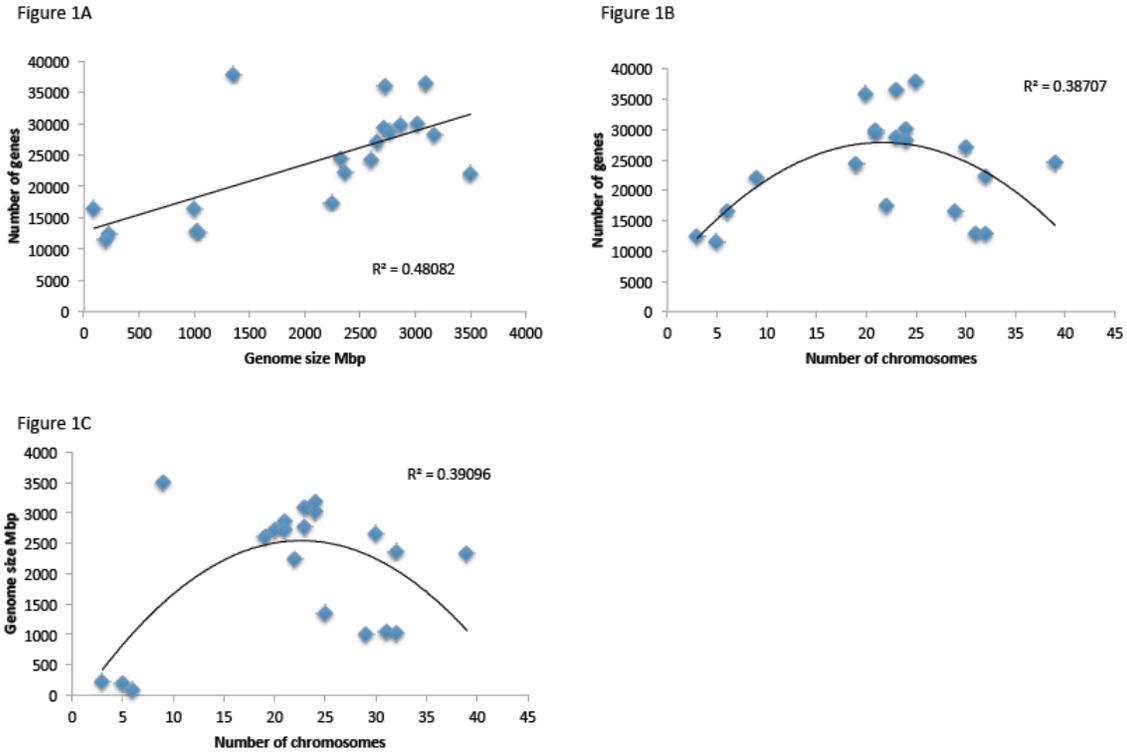

**Figure 2.** Chromosome size in base pairs (blue area) and gene complement (red area) per chromosome plotted out (log scale) for individual chromosomes in representative animals. Individual chromosome number given on X-axis. (A) *Homo sapiens* (humans); (B) *Gallus gallus* (Chicken); (C) *Anolis carolinensis* (Green anole, 6 chromosomes and 7 linkage groups); (D) *Oryctolagus cuniculus* (Rabbit).

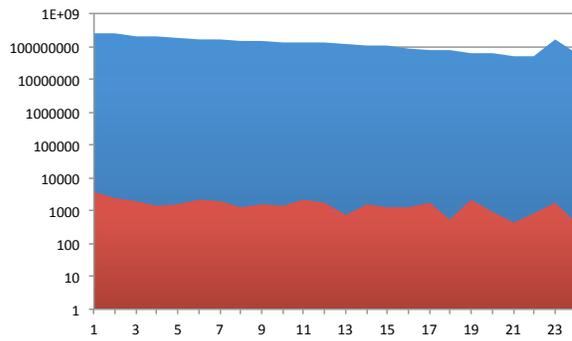

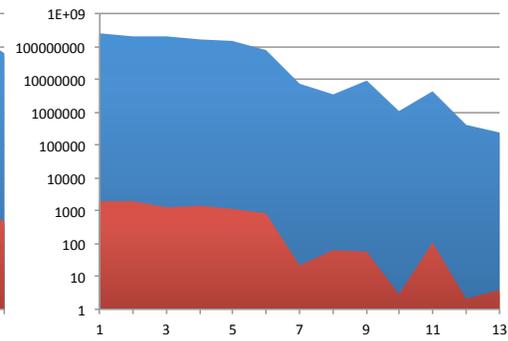

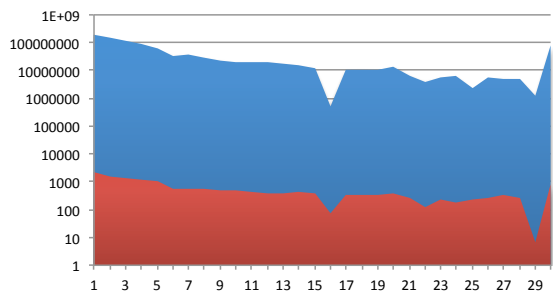

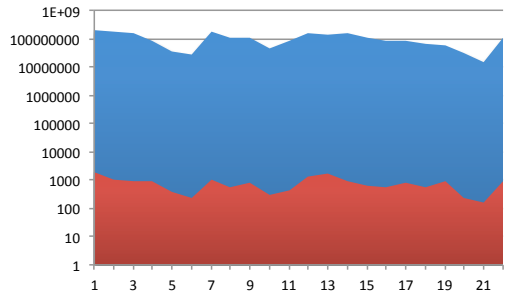

**Figure 3.** *Gene frequency* function [1] plotted against chromosome number in the animals examined (N=20). Blue rectangles depict average *gene* frequency for each species. Similar gene frequency per chromosome in all animals, with a slight decline as chromosomal complement of animal increases.

Figure 3

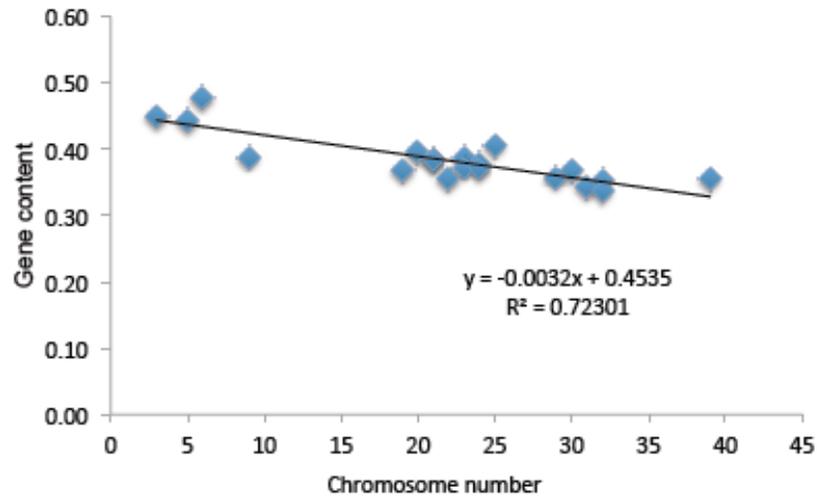

**Figure 4.** Average *gene distribution* across entire genome of different animals ordered by chromosomal complement in the animal. (A) Median *gene distribution function* [2] plotted against haploid chromosomal content (N=20), blue rectangles depict median gene distribution for each animal; (B) Median genes per chromosome plotted (blue rectangles) against haploid chromosomal content (N=20)

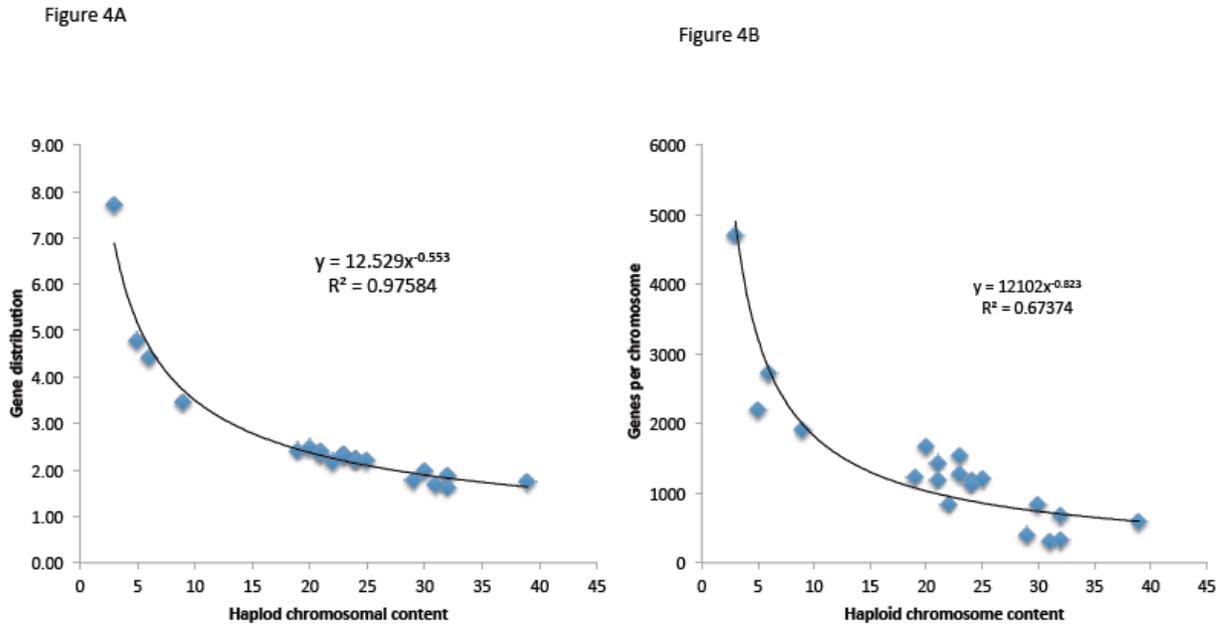

**Figure 5.** Power Law relationship in chromosome size and gene content. (B) *gene spacing* [3] and gene complement of each chromosome for humans. (C) *Gene spacing* ranked in order of magnitude, for individual chromosomes in all the genomes examined, demonstrates power distribution. Red squares, average gene spacing per chromosome ($R^2$= 0.58). Each data point corresponds to an individual chromosome.

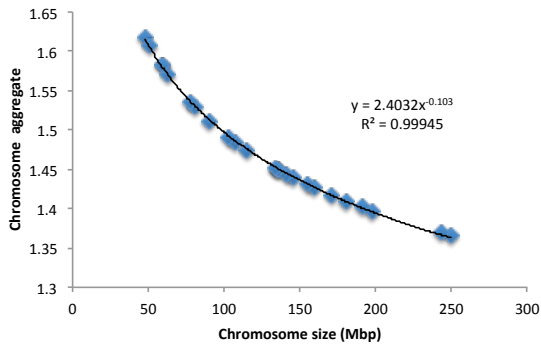

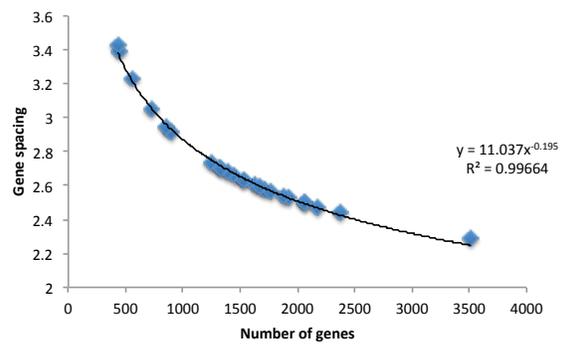

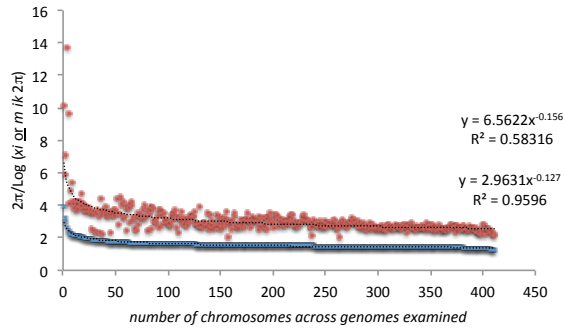

**Figure 6.** Median *gene frequency* [1] plotted against reciprocal of *gene spacing* [3] in different animals. (N=20) Linear decrease in gene frequency as the spacing between the genes increases in more evolved animals.

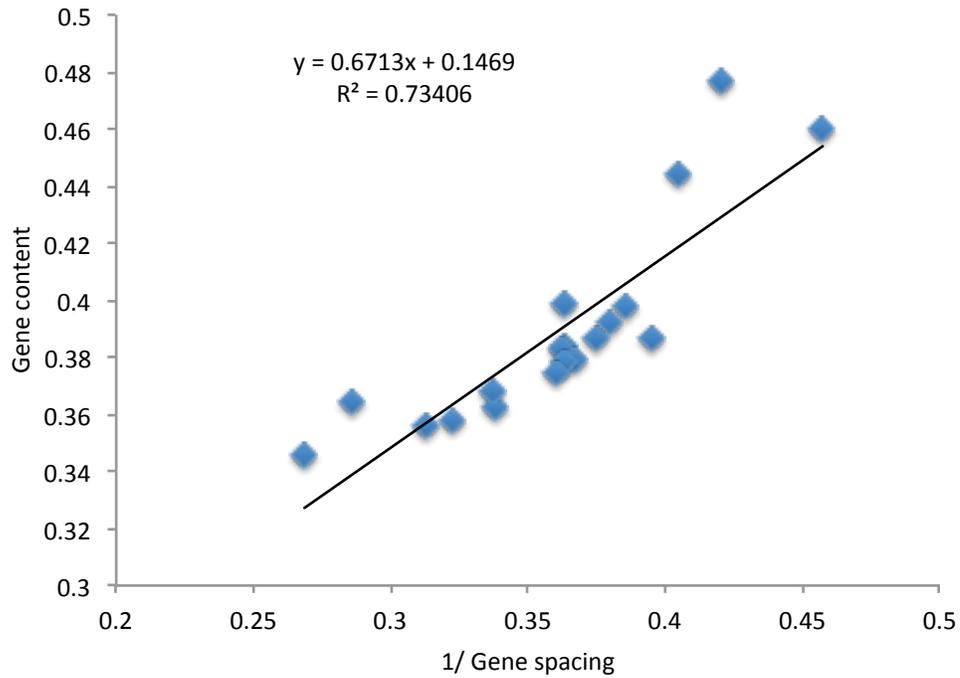

**Figure 7.** A model depicting the relationship between gene frequency, gene distribution and gene spacing on three hypothetical genomes with increasing chromosomal complement A, B and C. Two partial chromosomes from each of the genomes are shown. Gene content remians relatively unchanged, with proportional decline in gene frequency and gene distribution function and increase in gene spacing as animal proceed from simple to more complex organism from left to right. This implies that genes become more disperse in a mathematically ordered fashion as the chromosomal content of the animal increases.

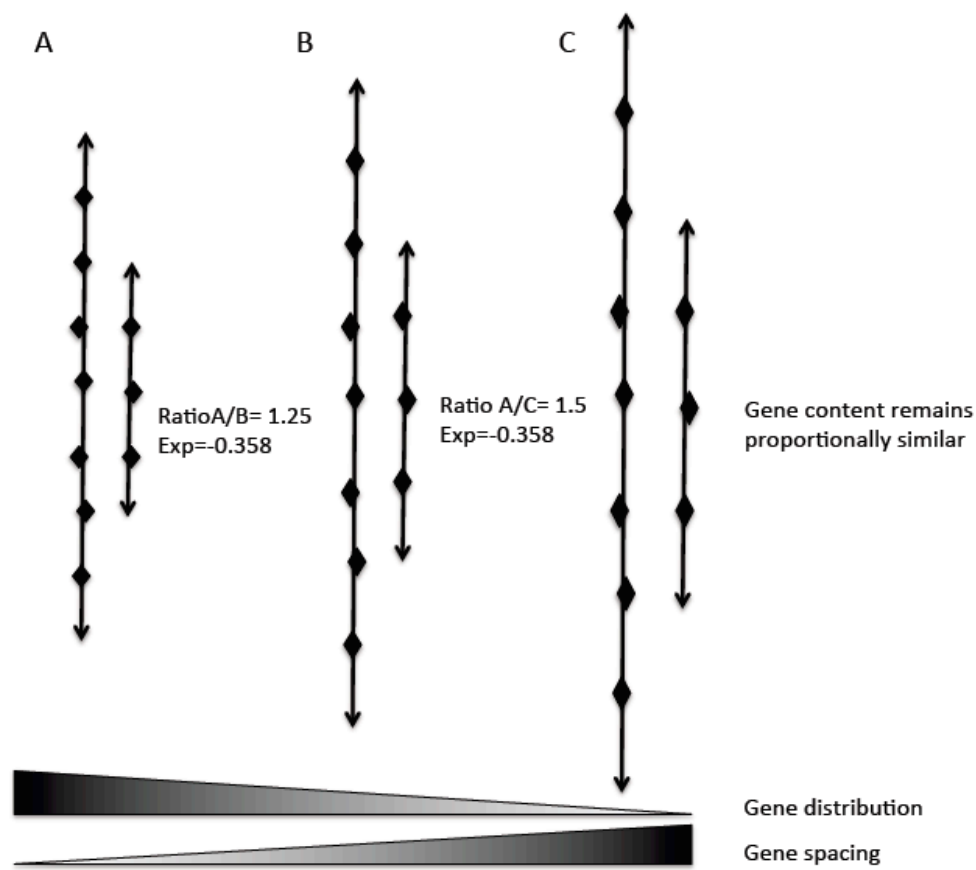

**Appendix A.** Riemann $\zeta$ function.

Given the observation that log-transformed quantitative relationships describing gene content of various chromosomes followed principles analogous to complex number iteration involving harmonic series, that is, self-similar fractals, it was hypothesized that gene distribution in chromosomes would be similar to the distribution of the non-trivial zeros of Riemann's $\zeta$ function. Therefore to further investigate the rules describing gene distribution on chromosomes, the relative relationships between the gene complement on each chromosome and their relative sizes was examined using the relationship, which determines the average spacing for non-trivial Riemann zeta-zeros. Zeta functions were first developed in number analytic theory to accurately determine the probability of finding the number of primes $P(N)$ in a sequence of real numbers, $N$. Carl Friedrich Guass determined that $P(N)$ was approximately equal to $N / \mathrm{Log}\ (N)$, though the relationship is associated with an error when compared with the true $P(N)$. While this error decreases as $N$ increases, its asymptotic limit is non-zero even for large $N$. Leonard Euler further refined this estimate with the development of the zeta function ($\zeta$) in his work on harmonic power series, $\zeta(s) = \Sigma\,n^{-s}$, where $n$ takes on values from 1 to $\infty$, and $s$ is substituted for $N$. Bernhard Riemann extended the zeta function to the complex number plane, calculating $\zeta$ for complex values of $s$ ($z = x + ti$), where $x$ represents the real component, (i.e., …, -2, -1, 0, 1, 2, …) on the 'real' $x$-axis in the complex plane, and $t$ is the real part of the 'imaginary' component, $ti$, with $i = \sqrt{-1}$, on the 'imaginary' $y$-axis in the complex plane. Application of the resulting correction factor in the earlier prime number estimates provides the best approximation of $P(N)$ to date. Riemann further determined that for complex numbers with a real, positive $x$ value of 0.5, $\zeta$ periodically takes on a value of zero as one goes up the 'imaginary' $y$-axis in the complex plane. These are termed, 'non-trivial' zeros of the $\zeta$ function, and in the complex plane are distributed on a 'critical line' extending up from 0.5 on the 'real' $x$-axis in the 'argument' ($s$) plane. Indeed when plotted in a complex 'value' plane – independently of $s$ – the value of $\zeta$ continually cycles in a clockwise manner through the complex plane, repeatedly returning to the origin as the value of $t$ rises in the imaginary ($ti$) component of the complex number argument ($s$). These 'non-trivial zeta zeros' are non-randomly distributed on the critical line, and become closer together as the height of the critical line increases, and $\zeta$ cycles through zero with increasing frequency in the 'value' plane. Andrew Odlyzko determined that the average space between the non-trivial zeros of the Riemann zeta function on the critical line is $\sim 2\pi / \mathrm{Log}\,(T / 2\pi)$, where $T$ represents the value of $t$ on the complex axis. (22) In this equation, which is analogous to the above relationship ($N / Log\,(N)$), $2\pi$ represents the 'period' of the periodic function ($f\,(y)=sin\,x$) exemplified by the sine wave, where $x$ is angle measured in radians. It implies that the spacing between successive zeros decreases logarithmically as one goes up the critical line.

**Appendix B.** Logarithms and self-similarity.

Phenomena described by large numbers are often difficult to comprehend and visualize, because every day life experiences tend to take on a more limited range of values. Large numbers are also cumbersome when used in mathematical operations and tend to overshadow smaller values when comparisons are made across different scales of measurement.  Logarithmic transformation of numbers may overcome this barrier to understanding relationship between phenomena, which encompass varying scales of measurement. Log-transformation of a *number* involves raising a specific *base* to an *exponent*, which then yields the *number*.  The exponent is called the *logarithm* of the *number*, for that *base.* As an example when the *base*, 10, is raised to the *exponent*, 2, it yields the *number*, 100; thus the *base-10 logarithm* of 100 is 2. An important property of logarithms is that they can be used to compare values across differing scales of measurement. As seen in Figure S1, when log-X is plotted against the value of X, these values maintain a similar distribution regardless of the scale of measurement, further, this holds true regardless of the base used to calculate the logarithms.  This property of logarithms is used to explore self-similarity in various systems across different scales of measurement.

**Figure S1.** Plots depicting the value of the natural-logarithm (*base*, *e*=2.718...), logarithm *base-5* and *base-10* plotted against the value of X for three different scales of measurement; 1 to 1000; 1000 to 1000000 and finally 1000000 to 1000000000.

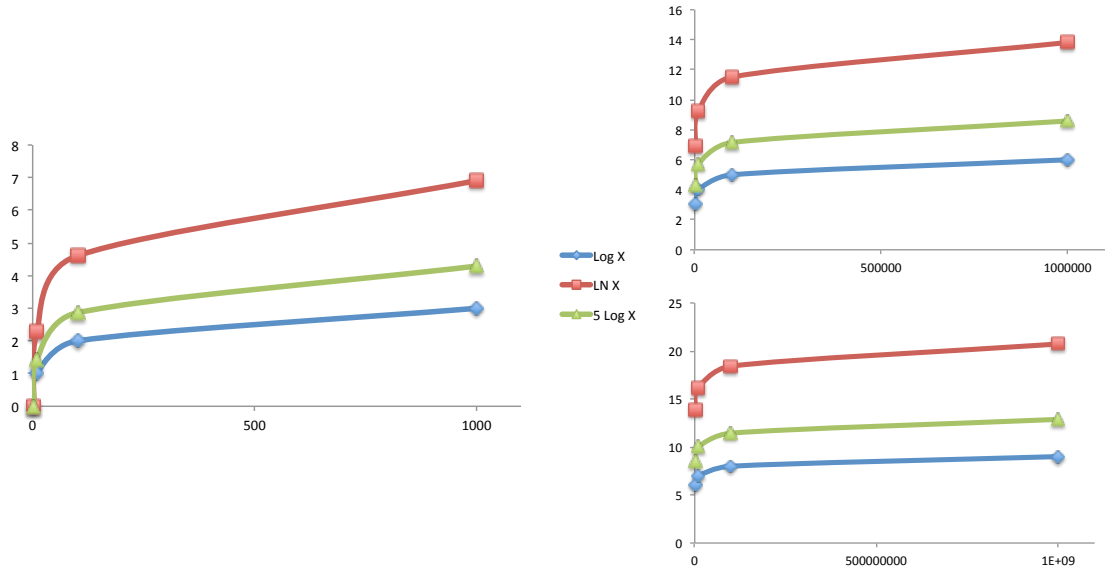

**Acknowledgements:** The authors gratefully acknowledge Dr. Roy Sabo from the VCU Department of Biostatistics for performing statistical analysis presented. AT was supported by research funding from the NIH-NCI Cancer Center Support Grant (P30-CA016059; PI: Gordon Ginder, MD).

**References.**